\documentclass[prl,notitlepage,citeautoscript,floatfix,amsmath,amssymb,superscriptaddress]{revtex4-2}%

\usepackage{graphicx}
\usepackage{dcolumn}
\usepackage{bm}
\usepackage{times}

\begin{document}


\title{Diffusion in a rough potential: Dual-scale structure and regime crossovers}
\author{Carlos E. Colosqui}
\email{carlos.colosqui@stonybrook.edu}
\affiliation{Department of Mechanical Engineering, Stony Brook University, Stony Brook, NY 11794, USA}
\affiliation{Department of Applied Mathematics \& Statistics, Stony Brook University, Stony Brook, NY 11794, USA}

%
%
%
%
%
\begin{abstract}
Diffusion in a ``rough'' potential parameterized by a reaction coordinate $q$ is relevant to a wide spectrum of problems ranging from protein folding and charge transport in complex media to colloidal stabilization and self-assembly.
This work studies the case of a potential having coarse-scale structure with characteristic energy barrier $\Delta U$ and period $\ell$, and fine-scale ``roughness'' of magnitude $\Delta U'\lesssim \Delta U$ and small period $\ell'\ll \ell$.
Numerical solution of the Smoluchowski equation and analytical predictions from Kramers theory document distinct regimes at different distances $|\Delta q|=|q-q_E|$ from stable equilibrium at $q=q_E$.
The physical diffusivity $D$ prescribed by dissipative effects can be observed farther than a distance $|\Delta q'| \propto (\Delta U'/\ell' + \Delta U/\ell)$.
Rescaling the physical diffusivity to account for the fine-scale ``roughness'' is strictly valid when 
$|\Delta q| < \Delta q_I \propto (\Delta U'/\ell' - \Delta U/\ell)$.
Farther than a critical distance $\Delta q_{II}\propto \Delta U/\ell$ the diffusion process is free of coarse-scale metastable states, which facilitates determining the effective diffusivity $D'$ from the reaction coordinate trajectory.
Closer to equilibrium the coarse-scale structure induces two diffusive regimes: nearly logarithmic evolution for $\Delta q_{II} > |\Delta q| > \Delta q_{III}$ and exponential decay over time for $|\Delta q| < \Delta q_{III}\propto 1/\ell$.
The effective diffusivity derived in this work is sensitive to the coarse- and fine-scale energy barriers and periods, and for $\ell'/\ell \to 0$ and $\Delta U'/k_B T \gg 1$ agrees closely with mean first-passage time estimates currently employed, which depend solely on the fine-scale energy barrier.
\end{abstract}
%
%
%
\maketitle

{\let\newpage\relax\maketitle}
\section{\label{sec:intro}Introduction}
The temporal evolution of reaction coordinates or collective variables in complex fluctuating systems commonly follows a drift-diffusion process in energy potentials with a coarse-scale structure that is experimentally observable and fine-scale ``roughness'' not readily observed.
Diffusion in a so-called ``rough'' potential  \cite{zwanzig1988} is relevant to numerous applications of current active research, such as protein folding \cite{zwanzig1995simple,dobson2003,gebhardt2010full,chung2015structural,mallamace2016energy,englander2017case,dill2017protein}, mass and charge transport in complex biological media and nanomaterials \cite{kalra2003,kim2013selective,di2013fast,lozada2016sieving,abraham2017,astumian2018stochastically,al2020large}, or stabilization and self-assembly of colloidal materials \cite{beltran2011free,hanes2012colloids,kraft2012surface,manoharan2015colloidal,zanini2017universal,liu2018capillary,hu2020spatiotemporal,srivastava2020dual}.
Low-dimensional potentials represent the projection of very complex multidimensional energy landscapes onto a small set of collective variables and can be densely populated by local minima over extremely fine scales that cannot be easily observed or resolved.
When such fine-scale minima are separated by energy barriers of magnitude larger than the thermal energy $k_B T$ they can give rise to metastable states that dramatically hinder the diffusion process, which results in the observation of an apparent or ``effective'' diffusivity $D'$ much smaller than the physical diffusivity.
Indeed, the physical diffusivity $D=k_B T/\xi$ determined by a damping coefficient $\xi$ satisfying dissipation-fluctuation relations \cite{kubo1966fluctuation} is only readily observed in the case of ``smooth'' potentials with negligible fine-scale energy barriers.

The observable coarse-scale structure of a potential, characterized by an energy barrier $\Delta U$ and period $\ell$, can be reconstructed with reasonable fidelity from sufficiently large sets of non-equilibrium reaction coordinate trajectories $q(t)$ by using different sampling techniques \cite{jarzynski1997,hummer2001free,gore2003bias,chipot2007free}. 
For a smooth potential $U_S(q)$ with a well-defined characteristic energy barrier magnitude and period and a known physical diffusivity, Kramers theory \cite{kramers1940} can be applied to predict accurately kinetic rates. 
Predicting or interpreting reaction coordinate trajectories and kinetic rates becomes significantly more difficult for a rough potential $U(q)=U_S+u'$ having a fine-scale perturbation or ``roughness'' $u'(q)$ with characteristic energy barrier magnitude $\Delta U'\lesssim \Delta U$ and very small periods $\ell' \ll \ell$.
A commonly adopted is based on the conjecture by R. Zwanzig \cite{zwanzig1988} that the physical diffusivity $D$ can be replaced by an effective diffusivity $D'$ in order to account for the fine-scale roughness of the potential; this is found to be valid under certain range of conditions specified in this work.
Furthermore, solving for the mean first-passage time (mfpt) leads to \cite{zwanzig1988} $D'=D \times f(\Delta U'/k_B T)$ with 
$f=\langle \exp(-u'/k_B T)\rangle\times \langle \exp(u'/k_B T)\rangle\le 1$; here, the brackets $\langle ~~ \rangle$ indicate spatial average along $q$. 
The effective diffusivity estimated by the mfpt has been often employed in protein folding to account for the presence of unobservable fine-scale ``roughness'' $u'$ with significant energy barriers $\Delta U'=$~3--7~$k_B T$ \cite{hyeon2003can,nevo2005direct,janovjak2007,wensley2010exp,yu2015protein}.
It is worth noticing that Zwanzig's conjecture \cite{zwanzig1988} amounts to rescaling time as function of the fine-scale energy barrier, regardless of the particular coarse-scale structure of the potential. 
According to this approach, diffusion in the ``rough'' potential $U=U_S+u'$ must result in reaction coordinate trajectories with the same functional form for that of the smooth potential $U_S$, albeit stretched over a larger time scale.  

Complex nonequilibrium behaviors with crossovers between different diffusive regimes have been reported in recent studies of Brownian motion in different colloidal systems due to the presence of energy barriers induced by nanoscale surface roughness or localized heterogeneities \cite{kaz2011,colosqui2013,colosqui2015,colosqui2016,keal2018colloidal,jose2018,zhao2023anomalous,singletary2024kinetic}.    
The trajectories of collective variables or reaction coordinates describing such systems show transitions between power-law, exponential, and logarithmic evolutions for different far-from-equilibrium conditions \cite{colosqui2016,rahmani2016,keal2018colloidal,jose2018,zhao2023anomalous,singletary2024kinetic}.
Although the fundamental physical process corresponds to diffusion in a rough potential \cite{zwanzig1988}, the different trajectories and regime crossovers observed cannot be readily accounted for by rescaling the physical time and diffusivity.
This work aims to elucidate the conditions under which it is valid to use an effective diffusivity to accurately model diffusion in ``rough'' potentials with coarse- and fine-scale energy barriers of arbitrary magnitude and period.  
The problem is revisited by performing numerical solution of the Smoluchowski equation and comparing against analytical predictions from Kramers theory for the case of a `rough'' harmonic potential comprising a coarse- and fine-scale structure with characteristic amplitudes $\Delta U\gtrsim \Delta U' >k_B T$ and periods $\ell' \ll \ell \le|q-q_E|$.
%
%
The analysis and results presented in this work can be particularly useful to interpret nonequilibrium reaction coordinate trajectories obtained from experimental techniques such as single-molecule optical and force spectrometry \cite{dudko2008theory,woodside2014,neupane2016protein} or colloidal probe atomic force microscopy \cite{ducker1991direct,ally2012,zanini2018}.

\section{\label{sec:model} Theory}
%
Given a fluctuating observable $q'(t)$ from a dynamic stochastic process such as Brownian motion, its expected value or noise-averaged trajectory
$q(t)=\langle q'\rangle=\int p(q',t) q' dq'$ can be predicted from the probability density function $p(q,t)$s.
The probability density function in multidimensional phase space is governed by Fokker-Planck type equations, which under proper simplifying assumptions \cite{kramers1940,hanggi1986,colosqui2015} can be reduced to a one-dimensional drift-diffusion equation
\begin{equation}
\frac{\partial}{\partial t}p(q,t)
=
\frac{\partial}{\partial q}
\left[D e^{-\frac{U}{k_BT}}\frac{\partial}{\partial q}e^{\frac{U}{k_BT}}p(q,t)\right]  
\label{eq:smoluchowski}
\end{equation}
that is commonly referred to as the Smoluchowski equation.
%
%
It is worth noticing that the Smoluchowski diffusion equation (Eq.~\ref{eq:smoluchowski}) is a mesoscopic deterministic description valid for uncorrelated thermal motion and overdamped systems where the energy potential and its derivatives do not vary significantly over the length scale $\ell_D= D/v_{rms}$ determined by the physical diffusivity $D$ and the characteristic (root-mean-square) velocity $v_{rms}$ of the thermal motion \cite{hanggi1986,colosqui2017}.

For the sake of analytical simplicity, and following the analysis in Ref.~\citenum{zwanzig1988}, this work considers the case of constant thermal energy $k_B T=\mathrm{const.}$ and physical diffusivity $D=\mathrm{const}$.
Moreover, the studied drift-diffusion process governed by Eq.~\ref{eq:smoluchowski} takes place in the 1D energy potential
\begin{equation}
U(q)=\frac{K}{2}(q-q_S)^2-F q
+\frac{\Delta U}{2} 
\cos\left(\frac{2\pi q}{\ell}+\varphi\right)
+u'(q)
\label{eq:potential}
\end{equation}
that comprises: (i) a ``smooth'' potential $U_S$ modeled as a harmonic well with curvature $K$ and a global minimum or native state at $q=q_S$, plus a single-mode perturbation of amplitude $\Delta U$ and period $\ell$; (ii) a linear term due to an applied force of magnitude $F$; and (iii) a fine-scale perturbation or ``roughness'' $u'$.
The equilibrium coordinate $q_E=q_S+F/K$ defines the stable equilibrium state observed for the coarse-scale potential under external forcing; accordingly, $\varphi=\pi (1-2q_E/\ell)$ is set to be the phase in Eq.~\ref{eq:potential}.
The fine-scale ``roughness'' in Eq.~\ref{eq:potential} takes the form of another single-mode perturbation
\begin{equation}
u'(q)=\frac{\Delta U'}{2} \cos\left(\frac{2\pi q}{\ell'}+\varphi'\right),
\label{eq:roughness}
\end{equation}
with small amplitude $\Delta U'\lesssim \Delta U$ and period $\ell' \ll \ell$.
The phase of the fine-scale roughness in Eq.~\ref{eq:roughness} produces no relevant effects for the case of very small periods $\ell' \ll |q-q_E|$; hereafter, $\varphi'=\pi(1-2\pi q_E/\ell')$ is chosen so that the global minimum of the rough potential lies exactly at $q_E$. 

\begin{figure*}[t]
\center
\includegraphics[angle=0,width=0.9\linewidth]{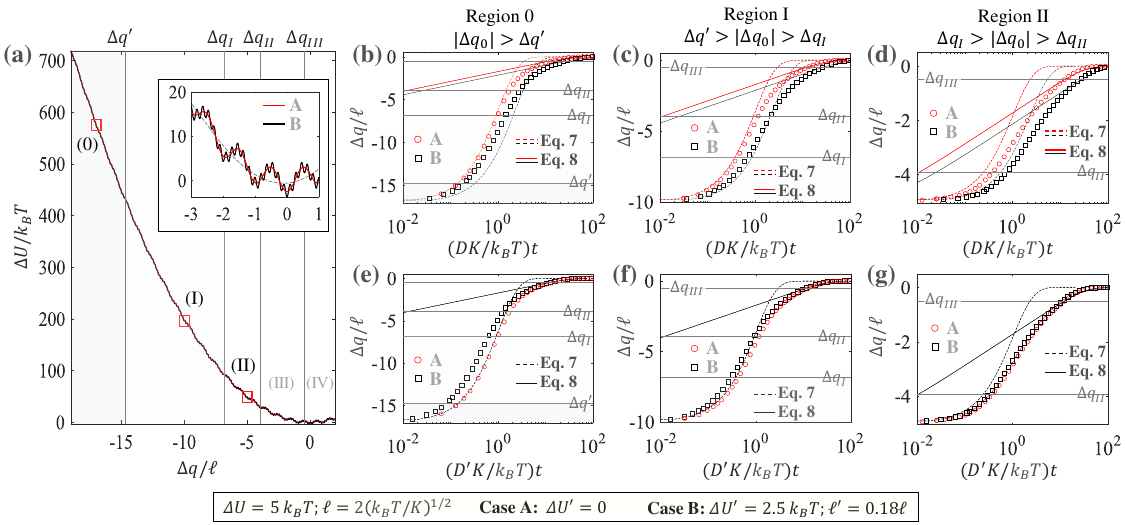}
\caption{\label{fig:1} Reaction coordinate trajectories for multiple coarse-scale metastable states and initial separations from equilibrium $|\Delta q_0|>\Delta q_{II}>\ell$. 
(a) Energy potential given in Eq.~\ref{eq:potential} for a coarse-scale energy barrier $\Delta U= 5 k_B T$ and period $\ell = 2(k_B T/K)^{1/2}$ with (Case A) no fine-scale perturbation $\Delta U'=0$, and (Case B) fine-scale energy barrier $\Delta U'=2.5 k_B T$ and period $\ell'=0.18 \ell$. 
Regions 0--IV are bounded by the critical separations from equilibrium defined in the text, square markers indicate that initial conditions in panels (b)--(g) lie within Regions 0--III.
(b)--(d) Dimensionless separation from equilibrium $\Delta q(t)/\ell=(q-q_E)/\ell$ vs. time nondimensionalized with the physical diffusivity $D$. 
(e)--(g) Dimensionless separation from equilibrium vs. time nondimensionalized with the effective diffusivity $D'$ (Eq.~\ref{eq:diffusion}) for Case B. 
Numerical solution of Eq.~\ref{eq:smoluchowski} (markers) for Cases A--B, see legends, are compared against analytical predictions from Eq.~\ref{eq:fastexp} (dashed lines) and Eq.~\ref{eq:deltaq} (solid lines).  
}
\end{figure*}

The reaction coordinate evolution can only have metastable equilibrium states if there are local minima $q_o$ where $dU(q_o)/dq=0$. 
For the potential in Eq.~\ref{eq:potential} local minima can only exist when the distance from equilibrium $|\Delta q|=|q-q_E|$ is smaller than a critical value 
$\Delta q' = \pi(\Delta U/\ell+\Delta U'/\ell')/K$.
The physical diffusivity $D$ will thus dictate the reaction coordinate evolution sufficiently far from equilibrium where $|\Delta q|> \Delta q'$ (i.e., Region 0 in Fig.~\ref{fig:1}a). 
For $|\Delta q|\lesssim \Delta q'$ the coarse-scale curvature can prevent the occurrence of some local minima induced by the fine-scale roughness when 
$K \Delta q+\pi \Delta U/\ell-\pi \Delta U'/\ell' > 0$.
Hence, metastable states with a regular period independent of the reaction coordinate variation can only emerge when the distance from equilibrium is smaller than
\begin{equation}
\Delta q_I=\frac{\pi}{K}
\left(\frac{\Delta U'}{\ell'}-\frac{\Delta U}{\ell}\right).
\label{eq:dqc1}
\end{equation}
Adopting a constant (i.e., coordinate-independent) effective diffusivity $D'$ to model the fine-scale roughness $u'$ is thus strictly valid for near-equilibrium trajectories where the separation from equilibrium $|\Delta q| \lesssim \Delta q_I$ is smaller than a critical value in Eq.~\ref{eq:dqc1} that is prescribed by both the coarse- and fine-scale energy barrier magnitude and period. 

For distances $|\Delta q| < \Delta q_I$ the fine-scale motion is mediated by thermally activated transitions between metastable states and thus can be described by a rate equation 
$dq/dt=\ell_{+}\Gamma_{+} - \ell_{-}\Gamma_{+}$ 
where $\ell_\pm$ and $\Gamma_\pm$ are the separation between metastable states and transition rates, respectively, in the forward/backward ($+/-$) direction.  
In what follows, it is convenient to introduce the characteristic curvature or stiffness $K'=K+2\pi^2\Delta U/\ell^2$ induced by the coarse-scale structure. 
Provided that the modeled roughness $u'$ has a small period $\ell' \lesssim \sqrt{k_B T/K'}$, one can assume $\ell_+=\ell_-\simeq \ell'$ and then invoke Kramers theory \cite{kramers1940} to obtain the transition rates   
$\Gamma_{\pm}=(D'/\ell'^2)\exp[\pm (q_E-q)/L']$, where the effective diffusivity
\begin{equation}
D'=D \frac{\sqrt{\phi^2-1}}{\pi} \frac{\ell'}{L'}
\exp\left(-\frac{\Delta U'}{k_B T}-\frac{\ell'}{4L'} \right)
\label{eq:diffusion}
\end{equation}
is determined by the characteristic length
$L'=2 k_B T/K' \ell'$ and the ratio $\phi=2\pi^2 \Delta U'/ K' \ell'^2$ between the coarse- and fine-scale curvatures.  
The effective diffusivity $D'=D\times f(K,\Delta U,\ell,\Delta U',\ell')$ in Eq.~\ref{eq:diffusion} is thus obtained by rescaling the physical diffusivity with a function of both the coarse- and fine-scale spatial structure of the potential.

The potential in Eq.~\ref{eq:potential} has local minima solely induced by the coarse-scale energy barriers $\Delta U$ when the distance from equilibrium is smaller than
\begin{equation}
\Delta q_{II}=|q_{II}-q_E|=\frac{\pi\Delta U}{K\ell}.
\label{eq:dqc2}
\end{equation}
For $\Delta q_I > |\Delta q|>\Delta q_{II}$, coarse-scale metastable states cannot exist, while and the constant diffusivity $D'$ in Eq.~\ref{eq:diffusion} can be employed to account for the fine-scale roughness.
Hence, the temporal evolution of the reaction coordinate corresponds to that observed in a smooth harmonic potential of curvature $K$ with a single minimum at $q_E$
\begin{equation}
q(t)-q_E=(q_0-q_E) \exp\left(-\frac{D'K}{k_BT}t\right),
\label{eq:fastexp}
\end{equation}
where $q_0\equiv q(t=0)$ is the initial condition.

As the system further approaches equilibrium and $|\Delta q|<\Delta q_{II}$, metastable states induced by the coarse-scale energy barrier $\Delta U$ cause the crossover to a pseudo-logarithmic regime \cite{colosqui2013} for which the ensemble-averaged reaction coordinate trajectory predicted via Kramers theory is
\begin{equation}
q(t)-q_E= 2L~\mathrm{atanh} \left[c \exp\left( -\frac{D' \kappa}{k_BT} t\right) \right].
\label{eq:deltaq}
\end{equation}
Here, $c=\tanh[(q_0-q_E)/2L]$ is an integration constant defined by the initial condition, $L=2 k_B T/K \ell$ is the characteristic length for the coarse-scale thermally activated motion, and 
\begin{equation}
\kappa = \frac{\sqrt{4(\pi/\ell)^4\Delta U^2-K^2}}{\pi} \frac{\ell}{L} 
\exp\left(-\frac{\Delta U}{k_B T}-\frac{\ell}{4L} \right),
\label{eq:diffusion2}
\end{equation}
is an effective stiffness prescribed by the coarse-scale structure of the potential.
For $|\Delta q|\gg L$, the expression in Eq.~\ref{eq:deltaq} predicts a logarithmic-in-time decay to equilibrium 
$q-q_E=L \log[D'\kappa t/2k_BT+d]$ with $d=\exp[(q_0-q_E)/L]$. 

For $|\Delta q| < L$ the expression in Eq.~\ref{eq:deltaq} predicts a new exponential decay 
\begin{equation}
q(t)-q_E=(q_0-q_E) \exp\left(-\frac{D'' K}{k_B T} t \right),
\label{eq:slowexp}
\end{equation} 
prescribed by a much smaller coarse-scale effective diffusivity $D''=D'\times \kappa/K$ that accounts for both coarse- and fine-scale perturbations to the harmonic potential in Eq.~\ref{eq:potential}.
Notably, the exponential trajectory in Eq.~\ref{eq:slowexp} has the same functional form expected for a smooth harmonic potential and emerges when the separation from the expected equilibrium at $q_E\simeq q_S+F/K$ is smaller than a critical value
\begin{equation}
\Delta q_{III}=|q_{III}-q_E|=\frac{2 k_B T}{K \ell}
\label{eq:dqc3}
\end{equation}
determined by the period of the coarse-scale energy barriers.

\section{\label{sec:results} Results}
\begin{figure*}[t!]
\center
\includegraphics[angle=0,width=0.9\linewidth]{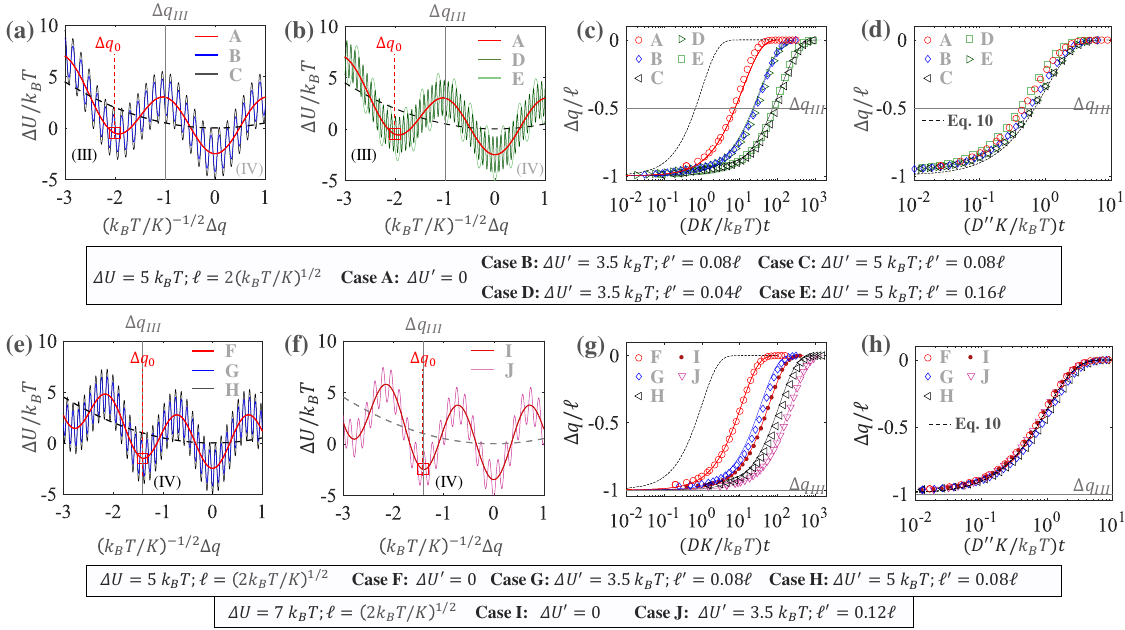}
\caption{\label{fig:2} Reaction coordinate trajectories for a single coarse-scale metastable state and initial separation from equilibrium $|\Delta q_0|=\ell<\Delta q_{II}$. 
(a)--(b) Energy potentials given in Eq.~\ref{eq:potential} for a coarse-scale energy barrier $\Delta U= 5 k_B T$ and period $\ell = 2(k_B T/K)^{1/2}$ with (Case A) no fine-scale perturbation $\Delta U'=0$, and (Cases B--E) fine-scale energy barriers $\Delta U'=$~3.5--5~$k_B T$ and periods $\ell'=$~0.04--0.16 $\ell$. 
Square markers indicate that the initial condition employed in panels (c)--(d) lies within Region III.
(c)--(d) Dimensionless separation from equilibrium $\Delta q(t)/\ell=(q-q_E)/\ell$ vs. time nondimensionalized with the physical diffusivity $D$ and coarse-scale effective diffusivity $D''=D'\times (\kappa/K)$ where $\kappa$ is given by Eq.~\ref{eq:diffusion2}. 
(e)--(f) Energy potentials for coarse-scale energy barriers $\Delta U=$~5--7~$k_B T$ and period $\ell = (2k_B T/K)^{1/2}$ with (Cases F \& I) no fine-scale perturbation $\Delta U'=0$, and (Cases G, H, J) fine-scale energy barriers $\Delta U'=$~3.5--5~$k_B T$ and periods $\ell'=$~0.08--0.16 $\ell$. 
Square markers indicate that the initial condition in panels (g)--(h) lies within Region IV.
(g)--(h) Dimensionless separation from equilibrium vs. time nondimensionalized with the physical diffusivity $D$ and coarse-scale diffusivity $D''$.   
Numerical solution of Eq.~\ref{eq:smoluchowski} (markers) for Cases A--J, see legends, are compared against analytical predictions from Eq.~\ref{eq:deltaq} (solid lines) and Eq.~\ref{eq:slowexp} (dashed lines).
}
\end{figure*}

%
Numerical solution of Eq.~~\ref{eq:smoluchowski} is performed using a conventional finite-difference procedure \cite{smith1985numerical,colosqui2015} with sufficiently high spatio-temporal resolution (i.e., uniform grid spacing $\Delta x \le 0.025 \ell'$ and time step $\Delta t \le 0.05 \Delta x^2/D$) to resolve accurately the drift-diffusion process over the fine-scale perturbation period $\ell'$.
The modeled initial condition  
$p(q,0)=(2\pi\sigma^2)^{-1/2} \exp[-(q-q_0)^2/2\sigma^2]$
is a narrow normal distribution with mean $q_0=q(0)$ and small standard deviation $\sigma=0.04\ell$. 
The boundary conditions employed are 
$\partial p(q_A,t)/\partial q =\partial p(q_B,t)/\partial q=0$, where $q_A=q_0-4\sqrt{k_B T/K}$ and $q_B=q_E+4\sqrt{k_B T/K}$ are the upper and lower bounds, respectively, of the simulation domain.
Reaction coordinate trajectories obtained from numerical solution of the 1D Smoluchowski diffusion equation (Eq.~\ref{eq:smoluchowski}) and the studied harmonic potential with dual-scale roughness (Eqs.~\ref{eq:potential}--\ref{eq:roughness}) are reported in Figs~\ref{fig:1}--\ref{fig:2} for different values of the coarse- and fine-scale energy barrier magnitude and period, curvature $K$ of the harmonic potential, and initial separations $\Delta q_0=q_0-q_E$ from equilibrium.
For the modeled system, varying the external force magnitude $F$ in Eq.~\ref{eq:potential} is equivalent to shifting the position of the expected equilibrium condition $q_E=q_S+F/K$.
Comparison of theoretical predictions from Eqs.~\ref{eq:diffusion}--\ref{eq:dqc3} against numerical solutions of the Smoluchowski equation confirms that qualitatively distinct trajectories $q(t)$ exist within different regions (0--III) bounded by the predicted critical separations from equilibrium $\Delta q_i$ (see Fig.~\ref{fig:1}a). 

In the region where $|\Delta q|=|q-q_E|>\Delta q'$ (Region 0) the reaction coordinate evolution (see Fig.~\ref{fig:1}b) follows closely the exponential decay for a smooth harmonic potential with curvature $K$ and determined by the physical diffusivity $D$.
In Region I where $|\Delta q|<\Delta q'$ (cf. Figs.~\ref{fig:1}b-c) the reaction coordinate trajectory $q(t)$ for $\Delta U'>0$ (Case B) deviates from the exponential decay prescribed by the physical diffusivity $D$ and gradually transitions to a slower exponential decay (Eq.~\ref{eq:fastexp}) prescribed by the effective diffusivity $D'$ predicted by Eq.~\ref{eq:diffusion}.
In region II where $\Delta q_I<|\Delta q|<\Delta q_{II}$ (cf. Figs.~\ref{fig:2}c-d) the reaction coordinate evolution for $\Delta U'>0$ completes the crossover to an exponential decay determined by the effective diffusivity $D'$.   

For the case of negligible fine-scale energy barrier $\Delta U' = 0$ (Case A in Figs.~\ref{fig:1}b-c) the reaction coordinate trajectories follow an exponential decay determined by the physical diffusivity $D$ when $|\Delta q|>q_{II}$ (Region I \& II).
As equilibrium is further approached (see Fig.~\ref{fig:2}d) and $|\Delta q|<\Delta q_{II}$ (Region III \& IV), the reaction coordinate trajectories for $\Delta U'\ge 0$ (Cases A \& B) gradually cross over from an exponential decay to a more complex pseudo-logarithmic evolution predicted by Eq.~\ref{eq:deltaq}.
As seen in Figs.\ref{fig:1}e-f, where the initial distance from equilibrium 
$|\Delta q_0|=|q_0-q_E|>\Delta q_I$ is larger than the critical value $\Delta q_I$ defined in Eq.~\ref{eq:dqc1}, rescaling time with an effective diffusivity $D'$ cannot collapse entirely the reaction coordinate trajectory for $\Delta U'>0$ into the trajectory for $\Delta U'=0$.
In this case, using an effective diffusivity $D'$ cannot produce an accurate description for diffusion in a potential with fine-scale roughness due to the crossover between trajectories determined by different diffusivity values that takes place in Regions I to II.
The collapse of the entire trajectory for $\Delta U'\ge 0$ into a single curve (see Fig.\ref{fig:1}g) is only observed for sufficiently small initial separation from equilibrium so that 
$|\Delta q|<\Delta q_I$ for $t\ge 0$.
Hence, only for $|\Delta q|<\Delta q_I$ one could employ a constant effective diffusivity $D'$ to account for the fine-scale roughness $u'$ over the entire reaction coordinate; despite the fact that $q(t)$ has a complex functional form that cannot fully described with a single analytical expression within Regions III and IV (cf. Fig.~\ref{fig:1}d \& Fig.~\ref{fig:1}g).

The results reported in Fig.~\ref{fig:2} correspond to 10 cases with different coarse- and fine-scale structure of the potential, for which diffusion takes place entirely in Regions III \& IV where the separation from equilibrium is always smaller than the critical value $\Delta q_{II}$ defined in Eq.~\ref{eq:dqc2}.
Reaction coordinate trajectories in Fig.~\ref{fig:2} therefore corresponds to conditions for which the effective diffusivity $D'$ in Eq.~\ref{eq:diffusion} can account for the fine-scale energy barriers $\Delta U'$ and the effective stiffness $\kappa$ in Eq.~\ref{eq:diffusion2} can be used to account for metastable states induced by the coarse-scale energy barriers $\Delta U$, which can be combined into a coarse-scale effective diffusivity $D''=D'\times \kappa/K$.  
As reported in Figs.~\ref{fig:2}a-c, reaction coordinate trajectories from numerical solution of the Smoluchowski equation (Cases A-D) can be closely predicted by the analytical expression in Eq.\ref{eq:deltaq} derived from Krammers theory.
The results in Fig.~\ref{fig:2}d for $|\Delta q_0|>\Delta \Delta q_{III}$ show that the reaction coordinate trajectories starting within Region IV cannot be fully collapsed into a single master curve when rescaling time by the coarse-scale effective diffusivity $D''=D'\times(\kappa/K)$ in order to account for both coarse- and fine-scale perturbations to the harmonic potential in Eq.~\ref{eq:potential}.
The trajectories for $|\Delta q_0|<\Delta q_{III}$ reported in Fig.~\ref{fig:2}e-g take place entirely within Region IV and follow closely a slow exponential decay, as predicted in Eq.~\ref{eq:slowexp}.    
As seen in Fig.~\ref{fig:2}h, the trajectories that lie entirely in Region IV, which is delimited by the critical separation $\Delta q_{III}\propto 1/\ell$ (Eq.~\ref{eq:dqc3}), can be entirely collapsed into the exponential trajectory (Eq.~\ref{eq:slowexp}) for a smooth harmonic potential when rescaling time with a coarse-scale diffusivity $D''$.

\section{\label{sec:conclusions}Conclusions and outlook}

The analysis in this work contributes to a better understanding of diffusion in a potential with coarse-scale structure that can be resolved experimentally and a fine-scale perturbation or ``roughness'' that is not directly observable but has significant effects on the time evolution of the studied reaction coordinate.   
Numerical solution of the 1D Smoluchowski diffusion equation and theoretical expressions derived from Kramers theory for the case of a harmonic potential with coarse- and fine-scale energy barriers document regimes with different functional forms of the ensemble-averaged trajectories of the reaction coordinate.
Explicit analytical predictions for the nonequilibrium trajectories $q(t)$ in each regime are derived in terms of the coarse- and fine-scale energy barrier magnitudes $\Delta U \gtrsim \Delta U' > 0$ and periods $\ell \gtrsim \ell'$, the curvature of the harmonic component or ``stiffness'' $K$, and the initial separation from the equilibrium condition $q_E=q_S+F/K$ with or without external force $F\ge 0$.
Moreover, critical separations from equilibrium for which regime crossovers occur are predicted by analytical expressions that yield close agreement with numerical solution of the Smoluchowski equation.
The emergence of trajectories $q(t)$ with different functional form restricts the use of an effective diffusivity obtained by rescaling the physical diffusivity, and should be expected for potentials with fine-scale roughness independently of the particular coarse-scale shape (e.g., linear-cubic) \cite{dudko2006,neupane2012}.

Zwanzig's conjecture \cite{zwanzig1988} on the use of an effective diffusivity to account for the (unobservable) fine-scale roughness of the potential is found to be strictly valid for separations from equilibrium smaller than a critical value $\Delta q_{I}$ (Eq.~\ref{eq:dqc1}) that is determined by the magnitude and period of the coarse- and fine-scale energy barriers.
Results reported in Figs.~\ref{fig:1}--\ref{fig:2} indicate that an effective diffusivity $D'$ (Eq.~\ref{eq:diffusion}) derived from Kramers theory can be used to predict reaction coordinate trajectories for potentials with different combinations of coarse- and fine-scale structure provided that 
$|q(t)-q_E|<\Delta q_I$ for $t\ge 0$.   
The effective diffusivity $D'$ in Eq.~\ref{eq:diffusion} is an estimate valid for the modeled sinusoidal roughness (Eq. ~\ref{eq:roughness}) in the case of small but finite periods $\ell'<\pi \Delta U/K' \ell$ and energy barriers $\Delta U'>\Delta U (\ell'/\ell)^2$.  
For the modeled sinusoidal roughness (Eq.~\ref{eq:roughness}) Zwanzig's approach based on solving the mfpt gives the effective diffusivity 
$D'=D I_0(\Delta U'/2 k_B T)^{-2}$ \cite{zwanzig1988} where $I_0$ is the modified Bessel function of the first-kind.
The effective diffusivity derived from the mfpt is solely determined by the fine-scale energy barrier magnitude $\Delta U'$ and gives values that are close to those predicted by Eq.~\ref{eq:diffusion} in the limit of vanishingly small periods $\ell'/\ell\to 0$ and large energy barriers $\Delta U'>3 k_B T$. 
The mfpt estimate, however, overpredicts the effective diffusivity value for the case of small but finite periods $\ell'\simeq$ 0.01--0.1, which can result in a significant overestimation of the fine-scale energy barrier $\Delta U'$.
Furthermore, whereas the effective diffusivity derived from the mfpt is always smaller than the physical diffusivity, the expression in Eq.~\ref{eq:diffusion} predicts that the effective diffusivity can be slightly larger than the physical diffusivity for a narrow range of moderate energy barriers $\Delta U'\simeq k_B T$, as reported in a previous work \cite{wagner1999}.

Given a set of experimental observations for the nonequilibrium trajectory of a collective variable or reaction coordinate $q$ (e.g., molecular extension of a protein, center-of-mass position of a nanoparticle), one could estimate the coarse-scale energy barrier $\Delta U$ and period $\ell$ using well-established sampling techniques \cite{jarzynski1997,hummer2001free,gore2003bias,chipot2007free} and employ analytical expressions in this work to determine the unresolved fine-scale energy barrier and periods, the effective diffusivity, and kinetic rates.
Alternatively, in the case that the studied regime crossovers can be experimentally observed, the expressions derived for the crossover coordinates and reaction coordinate trajectories can be employed to estimate both the coarse- and fine-scale energy barrier magnitudes and periods.
This could be particularly useful for interpreting data from single-molecule spectroscopy in constant-force mode \cite{woodside2006direct,gupta2011experimental} where regime crossovers could be observed by modulating the external force $F$.  

The theoretical analysis in this work corresponds to a model-dependent approach that can be readily extended to consider multi-mode decompositions and fine-scale Gaussian roughness.    
For overdamped systems the 1D Smoluchowski equation is an effective description suitable for developing ``model-free'' approaches to reconstruct low-dimensional energy landscapes from reaction coordinate trajectories.
An interesting possibility given the low computational cost of solving the 1D Smoluchowski equation is to integrate its numerical solution within an optimization algorithm for determining a large but finite number of modes composing the ``rough'' potential and the physical diffusivity that better describes the set of experimental observations. 
Such a ``model-free'' approach could provide a valuable alternative to statistical sampling methods currently employed to interpret nonequilibrium data for protein folding, colloidal assembly, mass/charge transport in nanomaterials, and other fluctuating systems.

\begin{acknowledgments}
The author acknowledges Joel Koplik and Jeffrey F. Morris for useful discussions.
This work was supported by the Office of Naval Research Contract \# N000141613178 and The National Science Foundation DMS-1614892.
\end{acknowledgments}


%

\end{document}